\documentclass[useAMS,usenatbib]{mn2e}
\usepackage{graphicx}
\usepackage{amsmath}
\usepackage{txfonts}
\usepackage{color}
\usepackage{ulem}
\usepackage{xstring}

\newcommand{\deleteline}[2]{\textsuperscript{\textcolor{black}{ #1}} \textcolor{black}{\sout{#2}}}

\def\lya{Ly$\alpha$}

\title[Ly$\alpha$ EW distribution at z $\sim$ 4.5]
{
Ly$\alpha$ Equivalent Width Distribution {\color{black} of Ly$\alpha$ Emitting Galaxies} at Redshift z $\sim$ 4.5
}

\author[Z. Y. Zheng,  et al. 2013]
{
Zhen-Ya Zheng$^{1, 2}$\thanks{E-mail: zhenya.zheng@asu.edu}, 
Jun-Xian Wang$^{2}$,
Sangeeta Malhotra$^{1}$,
James E. Rhoads$^{1}$, 
\newauthor
Steven L. Finkelstein$^{3}$, and
Keely Finkelstein$^{3}$.\\  
$^{1}$School of Earth and Space Exploration, Arizona State University, Tempe, AZ 85287\\
$^{2}$CAS Key laboratory for Research in Galaxies and Cosmology, Department of Astronomy, University of Science and Technology of China,\\\nonumber Hefei, Anhui 230026, China\\
$^{3}$Department of Astronomy, The University of Texas, Austin, TX 78712
}

\begin{document}

\date{Accepted XXXX. Received XXXX; in original form XXXX}

\pagerange{\pageref{firstpage}--\pageref{lastpage}} \pubyear{XXXX}

\maketitle

\label{firstpage}

\begin{abstract}
\lya\ line equivalent widths (EWs) provide important clues to the physical nature of high redshift 
Lyman alpha emitters (LAEs).
However, measuring the \lya\ EW distribution of high-z narrowband selected LAEs can be hard 
because many sources do not have well measured broadband photometry. We investigate  the 
possible biases in 
measuring  the intrinsic \lya\ EW distribution for a LAE sample at z $\sim$ 4.5 in the Extended Chandra Deep Field South (ECDFS).
We show that our source selection procedures produce only weak {\color{black} Eddington} 
type bias in both the intrinsic \lya\ luminosity function and the \lya\ EW distribution.  However, the observed EW distribution is severely biased if one only considers LAEs with detections in the continuum. Taking the broadband non-detections into account requires fitting the distribution of the broadband-to-narrowband ratio, which then gives a larger EW distribution scale length.  Assuming an exponential form of the intrinsic \lya\ EW distribution $d$N$/d$EW = N exp$^{-EW/W_0}$, we obtain W$_0$ = 167$^{+44}_{-19}$\AA\ (uncorrected for IGM absorption of \lya, {\color{black}and $\sigma_g$ = 160$^{+43}_{-12}$\AA\ for a gaussian EW distribution}).  We discuss the likely range of IGM absorption effects
in light of recent measurements of \lya\ line profiles and velocity offsets.
Our data are consistent with \lya\ EW being independent of UV luminosity (i.e., we do not see evidence for the ``Ando" effect).  Our simulations also imply that broad-band images should be 0.5-1 magnitude deeper than narrowband images for an effective and reasonably complete LAE survey.
Comparing with consistent measurements at other redshifts, we see a strong evolution in \lya\ EW distribution with redshift which goes as a power-law form of $W_0$ $\propto$ (1+z)$^{\xi}$, with $\xi$ = {\color{black}1.1$\pm$0.1 (0.6$\pm$0.1)} if no IGM corrections are applied to the \lya\ line; or $\xi$ = {\color{black}1.7$\pm$0.1 (1.2$\pm$0.1)} after applying a maximal IGM-absorption correction to \lya\ line {\color{black}for an exponential (a gaussian) EW distribution} from z = 0.3 to 6.5.
\end{abstract}


\section{INTRODUCTION}
\label{sec:intro}

Lyman alpha emission line galaxies (LAEs) are one of  two main classes of high-redshift star-forming galaxies selected by rest UV emission (the other class being Lyman Break Galaxies). With the recombination of hydrogen in the ambient interstellar medium (ISM), the ionizing radiation from young stars in galaxies generates prominent \lya\ emission.  Selected through a significant brightness excess in a narrowband image (where the \lya\ line is located) over a broadband image (which measures UV continuum), LAEs are typically younger, less massive, and less dusty than LBGs (Gawiser et al. 2007, Pirzkal et al. 2007, Finkelstein et al. 2009, Guaita et al. 2011). The measurement of the \lya\ line relative to the UV continuum 
level is defined as the equivalent width (EW = F$_{Ly\alpha}$/f$_{cont}$, where F$_{Ly\alpha}$ is the \lya\ line flux, and f$_{cont}$ is the UV continuum flux density).  Assuming a typical star formation history and initial mass function, dust-free galaxies with active star formation would have \lya\ EWs of 50-200\AA\ (Charlot \& Fall 1993). 

However, the observed EWs of LAE galaxies are often larger than expected at z $>$ 4. While stellar models predict a maximum intrinsic \lya\ EW of 240\AA, Malhotra \& Rhoads (2002) reported 60\% of the \lya\ emitters at z = 4.5 have intrinsic ("IGMcorr") EWs exceeding that value. This is also confirmed by Dawson et al. (2004, 2007), Wang et al. (2009), and Zheng et al. (2013) for larger LAE samples at z = 4.5, and  Shimasaku et al. (2006) and Ouchi et al. (2008) for LAE samples at z = 5.7. 
Possible explanations for these large EWs are very low metallicities (as galaxies undergo their first  throes of star formation, predicted by Partridge \& Peebles 1967), or enhancement of the Ly$\alpha$ EW via a clumpy ISM (Neufeld 1991, Hansen \& Oh 2006; Finkelstein et al.\ 2009), or some kind of Active Galactic Nucleus (AGN) contribution. However, recent studies have found evidence for dust in \lya\ galaxies (e.g., Finkelstein et al. 2008, 2009, Lai et al. 2007, Pirzkal et al. 2007), showing that \lya\ galaxies are not metal free, and thus large EWs are not generally due to primitive star formation.  Dust could produce weird radiative transfer effects and so allow \lya\ photons out, at least in some objects.  A large fraction of AGNs in the high-redshift LAE samples are also ruled out (Malhotra et al. 2003, Wang et al. 2004, and Zheng et al. 2010).

Malhotra et al. (2012) compared the UV size and star-formation intensity (i.e., UV luminosity per unit area) of LAEs and LBGs over redshift 2.25 $<$ z $<$ 6. They found that \lya-selected galaxies have a characteristic, constant, small size in rest-frame UV light, unlike LBGs which have been previously shown to decrease in linear size as $H(z)^{-1}$ with increasing redshift, and both LAEs and LBGs have a characteristic star-formation intensity. 
Thus evolution in physical properties of ISM in LAEs over redshifts could yield evolution in \lya\ EW distribution.

Measuring the intrinsic \lya\ EW distribution can be challenging. One reason is that many LAEs are not detected in broadband images, thus their \lya\ EW can't be well constrained. Another issue is that LAE selection criteria may have introduce selection biases  to the \lya\ EW distribution, which have to be carefully explored.
In this paper we run Monte-Carlo simulations to probe the intrinsic \lya\ EW distribution of a LAE sample at  $z$ $\sim$ 4.5 selected in 
the Extended Chandra Deep Field South (ECDFS) over a 0.34 deg$^2$ region.
The photometric surveys of this sample were presented by Finkelstein et al. (2008, 2009), and their spectroscopic followup and \lya\ luminosity function were presented by Zheng et al. (2013, hereafter Paper {\sc I}). 
We briefly introduce our photometric and spectroscopic observations in \S 2, then present the observed \lya\ equivalent width distributions in \S 3. We introduce the Monte Carlo simulations in \S 4, finally, discuss the simulation results and present the evolution of EW distribution over redshift range of 0.3--6.5 in \S 5. Throughout this work, we assume a cosmology  with $H_0$ = 70 km s$^{-1}$ Mpc$^{-1}$, $\Omega_m$ = 0.27, and $\Omega_\Lambda$ = 0.73 (Komatsu et al. 2011). At redshift $z$ = 4.5, the age of the universe was 1.38 Gyr, with a scale of 6.8 kpc/\arcsec, and a redshift change of $\delta$z = 0.03 implies a comoving distance change of 19.0 Mpc. Magnitudes are given in the $AB$ system.

\section{Data}
\label{sec:data}

\subsection{Photometric Candidates}
\label{sec:data:photoobs}
We have selected 112 LAE candidates at z $\sim$ 4.5 in the GOODS {\it Chandra} Deep Field South region (CDF-S; RA~03:31:54.02, Dec~$-27$:48:31.5)  in three narrow bands,  including 4 LAEs in NB656 (Finkelstein et al. 2008),  33  in NB665 and 75 in NB673 (Finkelstein et al. 2009). All the narrow band images were obtained with the MOSAIC II camera on CTIO Blanco 4 meter telescope. The NB665 and NB673 candidates were selected from the overlap region between the MOSAIC image and the ESO Imaging Survey (EIS, Arnouts et al. 2001). The NB656 candidates, however, were selected in a much smaller
area (the overlap region between the MOSAIC image and the GOODS CDF-S data), thus only four objects were selected.

The LAE selection criteria were introduced in Rhoads et al. (2000), Rhoads \& Malhotra (2001), and Finkelstein et al. (2009), which require a 5 $\sigma$ significance detection in the narrowband, a 4 $\sigma$ significance narrowband flux density excess over the R band, a factor of 2 ratio of narrowband flux to broadband flux density, and no more than 2 $\sigma$ detection in the B-band. Candidates with GOODS B-band coverage were further examined in the GOODS B-band image, and those with significant GOODS B-band detections were 
excluded (see Paper {\sc I} for details). The first three criteria ensure a significant line detection, while the last criterion checks that it is at z $>$ 4. The factor of 2 ratio of narrowband flux to broadband flux density ensure all candidates have EW$^{no-IGM-corr-on-Ly\alpha}_{rest}$ $>$ 9.0 \AA\ (see \S3). The broadband EIS R band data has a 5 $\sigma$ 
limit of m(R)$_{lim}$ = 25.6 measured within a 2\arcsec\ diameter aperture. The narrowband images are calibrated to the R band image. The 5 $\sigma$ magnitude limits of the  narrowband images (NB665 and NB673)  of m(NB)$_{lim}$ = 25.0 ensure detections of pure emission line flux $>$ 1.8 $\times$ 10$^{-17}$ ergs cm$^{-2}$ s$^{-1}$.

\subsection{Spectroscopic Observations}
\label{sec:data:specobs}

The spectroscopic followup were presented in Paper {\sc I}. The spectroscopic observations were taken with IMACS on Magellan Baade telescope.
We obtained spectra of 64 out of 112 LAE candidates (3 of 4 in NB656, 17 of 33 in NB665, and 44 of 75 in NB673), and 46 LAEs were spectroscopically confirmed as z $\sim$ 4.5 LAEs (3 in NB656, 11 in NB665, and 32 in NB673).  {\color{black} We did not find any emission line at the  wavelength region of the corresponding narrowband for the remaining 18 candidates, and none of them is an interloper. }
Due to the large uncertainties in the flux calibration of the spectroscopic data, in this paper we adopt \lya\ line flux and equivalent width from photometric data. Note that all targets with photometric line flux f$_{Ly\alpha}$ $>$ 3.7$\times$10$^{-17}$ ergs cm$^{-2}$ s$^{-1}$ are confirmed (see Figure 1 \& 2 in Paper {\sc I}), and targets with large EWs are confirmed at a significant higher fraction (see Figure \ref{ewr} and \ref{ewrlow}).


\section{Observed Ly$\alpha$ Equivalent Width Distribution}
\label{sec:results:obsew}
As pointed out by Shimasaku et al. (2006), the rest-frame EWs of LAEs from photometry are calculated either using narrowband and a non-overlapping broadband at redder wavelengths (e.g., Ouchi et al. 2008 for LAEs at z = 3.1, 3.7, and 5.7), or narrowband and an overlapping broadband (e.g., this work for LAEs at z = 4.5, and Ouchi et al. 2010 and Kashikawa et al. 2011 for LAEs at z = 6.5).
t
When using an overlapping broadband, an IGM transmission correction should be applied to the continuum. 
For our data, R band is $\sim$1-2 mag deeper than I band, so we choose R band and narrowband to measure the EWs for our z $\sim$ 4.5 LAEs.

Following an approach similar to Malhotra \& Rhoads (2002), we calculate the rest frame \lya\ equivalent widths of the LAEs. We use the relations 
\begin{eqnarray}
\frac{N}{W_N} = a_N \times  \frac{F_{Ly\alpha}}{W_N}  + b_N\times \frac{F_{Ly\alpha}}{EW_{rest} \times (1+z)} \\
\frac{R}{W_R}  =  a_R \times \frac{F_{Ly\alpha}}{W_R}  + b_R \times  \frac{F_{Ly\alpha}}{EW_{rest} \times (1+z)}. \\ \nonumber
\end{eqnarray}
Here $R$ and $N$ are the integrated fluxes in the broad R filter and narrowband filter.  $W_N$ is the
narrowband filter width, defined as $W_N = \int T_N d\lambda / \max(T_N)$, and $W_R$ is the
corresponding quantity for R band.  
F$_{Ly\alpha}$ is the observed \lya\ line flux, and F$_{Ly\alpha}$/(EW$_{rest}$ $\times$ (1+z)) is the continuum flux density redward of the \lya\ line.
The coefficients $b_R$ and $b_N$ account for IGM absorption (Madau 1995) of continuum emission in R and narrowband, respectively, assuming the \lya\ line 
sits in the center of the corresponding narrowband filter.
$a_N$ and $a_R$ correct the effect that the narrowband and R band filters are not top-hat and have considerably smaller transmission fraction (comparing with the peak of the transmission curve) at the wavelengths of the narrowband center (an effect ignored in Malhotra \& Rhoads 2002). Solving for EW, 
we obtain:
\begin{eqnarray}
EW_{rest} 	&  =   &  \frac{b_R\times N\times W_R -b_N\times R\times W_N}{a_N\times R-a_R\times N}\times \frac{1}{1+z}, 
\end{eqnarray}
For our $z\approx 4.5$ \lya\ search, the coefficients become
$a_N$ $\sim$ 1, $b_N$ $\sim$ 0.66, $a_R$ $\sim$ [0.82, 0.77, 0.74] and $b_R$ $\sim$ [0.63, 0.61, 0.59] assuming a composite LAE spectrum (line + continuum f$_\lambda(Cont.)$ $\propto$ $\lambda^{-2}$)  for the narrowband filters [NB656, NB665, NB673] respectively. 
Unlike Malhotra \& Rhoads (2002), here we make no correction to IGM absorption to the \lya\ line, which is still poorly understood (see \S\ref{sec:diss:evolution} below for further discussion).
Throughout this paper, if not specifically stated,  we present only \lya\ line EW before correction for IGM absorption to the \lya\ line.

\begin{figure}
\begin{center}
\includegraphics[totalheight=0.3\textheight]{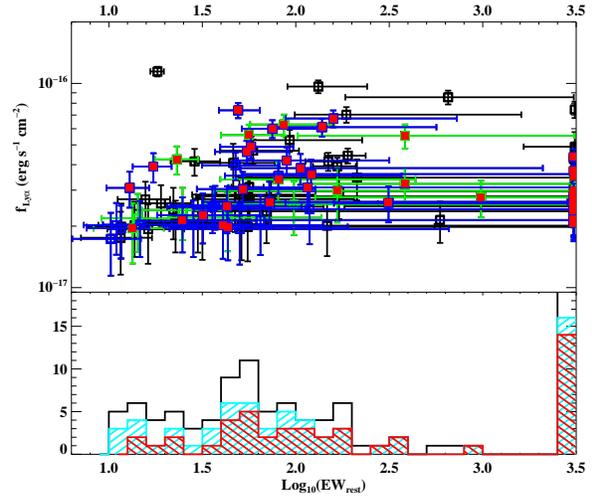}
\caption{Upper: The logarithmic EWs' distribution as a function of \lya\ fluxes for all candidates (empty black squares), targeted candidates (empty green and blue squares are targets in NB665 and NB673 images, respectively), and the spectroscopically confirmed LAEs (red filled squares) at z=4.5. Lower: The histogram distribution of log EWs for all candidates (black empty histogram), targets (cyan line-filled histogram), and confirmed LAEs (red line-filled histogram). } 
\label{ewr}
\end{center}
\end{figure}

In the upper panel of Figure \ref{ewr} we first plot observed \lya\ line flux versus line EW (from equation 3) for our LAE sample. Uncertainties in EWs were obtained through simulations by adding gaussian noise to broad-band and narrow-band flux densities. A clear trend can be seen that sources with larger EW also have larger uncertainties in EW. This is simply because of the much poorer constraints on continuum fluxes for larger EW sources (most of them have very weak continuum radiation). Specifically, calculations based on equation 1 produces negative continuum fluxes and thus negative EWs for some sources, which could be attributed to the large noise fluctuations in the broadband photometry. These sources with ``negative" line EW indeed have rather large line EW. In figure 1 we plot them at the right end by setting EW = 10$^{3.5}$ \AA\ (for display only). In the lower panel of Figure \ref{ewr}, we plot the EW histogram distribution for all candidates, targeted, and confirmed LAEs. We find that about 44\% (39\%) of the confirmed (candidate) LAEs show EW$_{rest}$ $>$ 154 \AA\ (intrinsic EW $>$ 240\AA, if applying a correction factor of 0.65 for IGM-correction on \lya\ line at z$\sim$4.5 assuming no velocity shift of the line), slightly lower than but consistent with the fraction of 50-60\% for the z = 4.5 LAEs in LALA fields (Malhotra \& Rhoads 2002, Wang et al. 2009). Dawson et al. (2007) measure the \lya\ EW from the spectra according to EW\'\ = $(F_l/f_{\lambda,r})/(1+z)$, where $F_l$ is the flux in the emission line and $f_{\lambda,r}$ is the measured red-side continuum flux density. 
The fraction of LAEs in Dawson et al. (2007) with EW greater than 
154\AA\  is 31$\pm$11\%, consistent with our results. 
 
\begin{figure}
\begin{center}
\includegraphics[totalheight=0.25\textheight]{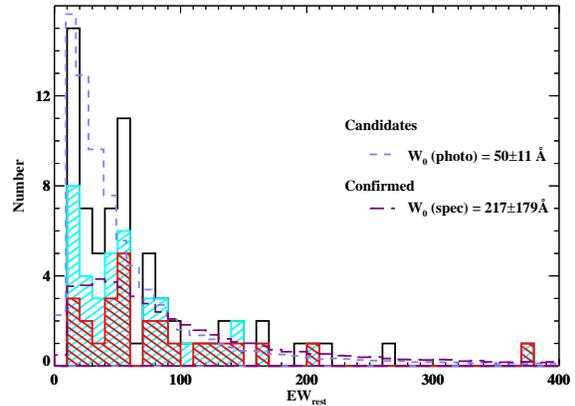}  
\caption{The \lya\ line EW distributions of LAEs with rest-frame EWs less than 400 \AA\ for all candidates. All photometric candidates, targets and confirmed LAEs are marked as {\color{black}empty black}, {\color{black}light-blue hatched} and red {\color{black}hatched} histograms. The EW distribution can be fitted with an exponential form $dN/d$EW = N exp$^{-EW/W_0}$, here we get $W_0$ = {\color{black}50}$\pm$11 \AA\ (blue dashed line) for all photometric candidates ({\color{black}empty black hatched} histogram, $\chi^2$/dof = {\color{black}21.1/18}) and $W_0$ = {\color{black}217$\pm$179} \AA\ (purple dashed line) for spectroscopically confirmed LAEs ({\color{black}red hatched} histogram, $\chi^2$/dof = {\color{black}4.7}/13). } 
\label{ewrlow}
\end{center}
\end{figure}

Figure \ref{ewrlow} shows the distributions of EW$_{rest}$ in the lower EW range (76 of 112 LAEs with EW $<$ 400\AA\ here, and their uncertainties in EWs are reasonably small, see figure \ref{ewr}). 
The EW distribution are often 
fitted with an exponential law of $d$N$/d$EW = N exp$^{-EW/W_0}$ or a positive gaussian distribution of $d$N$/d$EW = N $\frac{1}{\sqrt{2\pi \sigma_g^2}}  e^{-x^2/(2\sigma_g^2)}$ (Gronwall et al. 2007, Guaita et al. 2010, Nilsson et al. 2009). Here our photometric sample has a best-fit exponential scale of $W_0$ = {\color{black}50}$^{+11}_{-11}$ \AA. However, the $W_0$  is not well constrained for our spectroscopically confirmed sample ($W_{0,\  spec}$ = {\color{black}217$\pm$179} \AA), likely due to the smaller sample size and/or incompleteness in spectroscopic identifications of LAEs with low EWs.  
Assuming a Gaussian distribution, we obtain a $\sigma_{g}$ = {{\color{black}76$^{+11}_{-11}$} \AA\ for the photometric sample, while $\sigma_{g,\ spec}$ is also poorly constrained at $\sim$ {\color{black}189$^{+84}_{-84}$}\AA.

However, an exponential EW distribution with  $W_0$ = {\color{black}50}$^{+11}_{-11}$ \AA\ (or a gaussian EW distribution with $\sigma_{g}$  = {\color{black}76$^{+11}_{-11}$}\AA) implies that only $\sim$ 8\% (3\%) of sources with EW greater than 9.0 \AA\ have EW $>$ 154 \AA, apparently in contradiction to the fact that 39\% of our candidate LAEs have EW$_{rest}$ $>$ 154 \AA. This is simply because during the fitting to EW distribution, we excluded sources with ``negative" line EW,  and sources with photometric EW $>$ 400 \AA\ for which the uncertainty in EW is very large. Thus the $W_0$ or $\sigma_{g}$ from photometric sample was significantly under-estimated. In addition, the observed line EW distribution, especially at the low EW range, is likely sensitive to candidate selection criteria, and to the depth of R and narrow band images, thus could have been biased. For instance, a deeper R band image would allow more LAE candidates with smaller EW pass our selection. 
To confront these issues, below we do Monte Carlo simulations to obtain the intrinsic EW distribution in our LAE sample.

\section{Intrinsic \lya\ EW Distribution through Monte-Carlo Simulations}
\label{sec:results:ewsimu}

In order to obtain the intrinsic \lya\ line EW distribution from our LAE sample, we develop a Monte-Carlo approach to simulate the LAE selection processes described in Sec. 2.1. {\color{black}Our method is similar to that done by Shimasaku et al. (2006), however, their aim is to obtain the \lya\ luminosity function at z=5.7. Recent observations show that L$^*$ from observed \lya\ luminosity functions does not evolve significantly over the redshift range of 3 $<$ z $\leq$ 6.5 (see figure 15 of Paper {\sc I}), so we fix the L$^*$ in our simulation and check the selection process with variable EW distribution. We further use broadband to narrowband ratio instead of EW distribution in  fitting the intrinsic EW distribution. The band-ratio distribution has much better behaved errors than the EW distribution, and is less sensitive to objects  with low-EW values that are boosted above our selection threshold by  noise fluctuations. }  

Starting from an intrinsic \lya\ luminosity function, we build large artificial LAE samples by assigning \lya\ line luminosity to each source (L$_{Ly\alpha}$ range: 10$^{41.5}$ ergs/s $\leq$ L$_{Ly\alpha}$ $\leq$ 10$^{43.45}$ ergs/s, note the L$_{Ly\alpha}$ range of our real LAE sample is  [42.6,  43.3]).  Assuming their \lya\ line EW follows the exponential law $d$N$/d$EW = N $exp^{-EW/W_0}$ independent of \lya\ luminosity, we could further assign line EW to artificial sources (EW range: EW$\geq$ 1\AA, considerably below the EW limit of the real sample 9.0 \AA) and calculate their expected narrowband and R band fluxes. By adding Gaussian noise to the narrowband and R band fluxes (with noise level derived from our real data), and applying the same LAE selection criteria we adopted to select real LAE candidates, we obtain artificial LAE samples for various $W_0$ to compare with our real sample. 
We start our simulations by adopting a Ly$\alpha$ luminosity function following a Schechter Function of 
\begin{equation}
\Phi(L)d L  =  \frac{\Phi^*}{L^*} \left (\frac{L}{L^*}\right )^\alpha \exp\left (-\frac{L}{L^*}\right ) d L,
\end{equation}
with log$_{10}$( L$_*$)  = 42.75 and $\alpha$ = -1.5, which are the best-fit Ly$\alpha$ luminosity function parameters for our z = 4.5 LAE sample in Paper {\sc I}. 
Here we assign 3,200,000 simulated LAEs with \lya\ luminosity in the range of log$_{10}(L_{Ly\alpha})$ = [41.5, 43.45] (binsize = 0.03, and the brightest \lya\ luminosity bin has number $>$ 100). However, we note that the intrinsic luminosity function could be different from the observed one due to selection effects, and such differences should be measured with our simulations.  

In Figure \ref{ewfracevolve} we compare the input samples of the simulations with the output artificial samples by applying our selection criteria with intrinsic EW distribution in exponential \deleteline{}{and gaussian form, respectively} form {\color{black}(the gaussian form has similar patterns)}. As introduced in \S2.1,
the selection criteria (see also Finkelstein et al. 2009) are 5 $\sigma$ detection in narrowband (CR1: NB $\geq$ 5 $\sigma_{NB}$), a factor of 2 of narrowband over broad band (CR2: f$_{NB}$ $\geq$ 2 $\times$ f$_{R}$), a 4 $\sigma$ significance of narrowband over broad band (CR3: NB - R $\geq$ 4 $\times$ $\sqrt{\sigma_{NB}^2 + \sigma_{R}^2}$), and no more than 2 $\sigma$ detection in the B-band.



\begin{figure*}
\begin{center}
\includegraphics[totalheight=0.52\textheight]{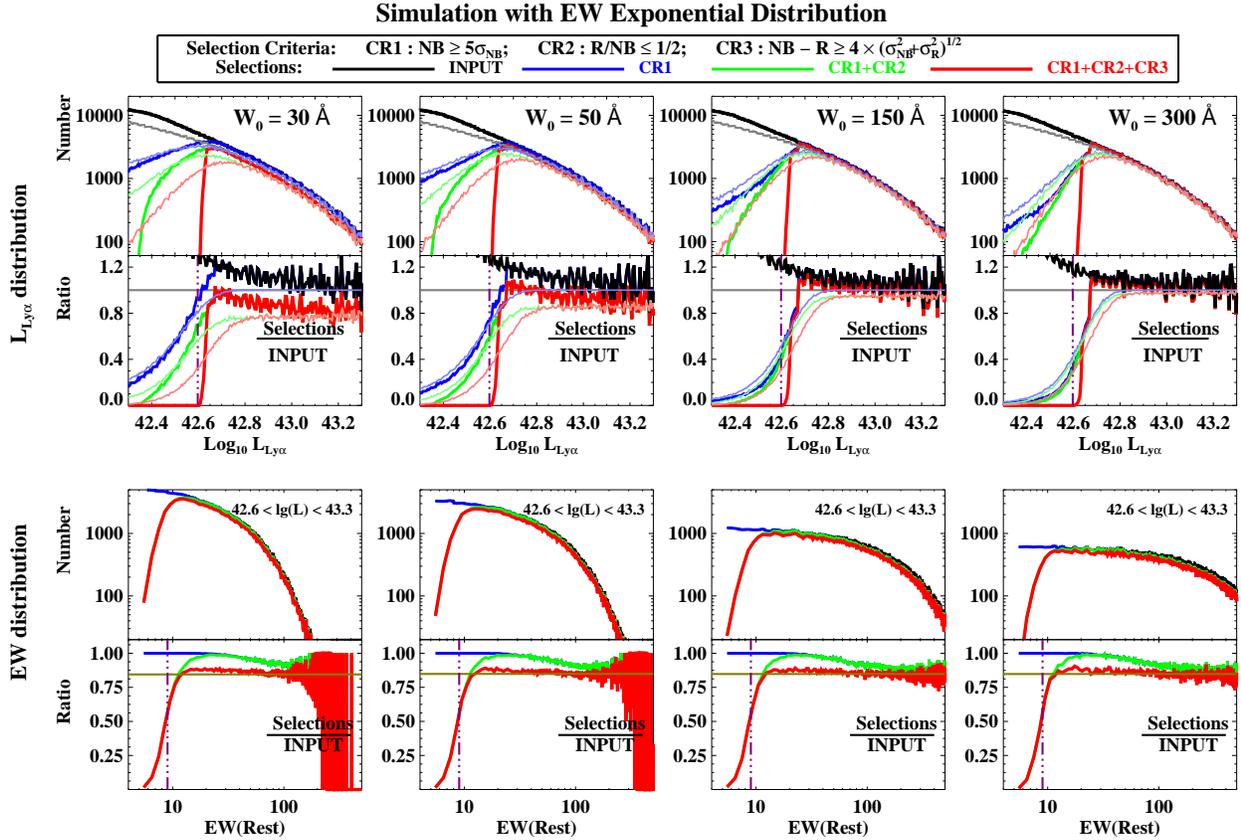}  
\caption{Monte-Carlo simulations on the selection process for different intrinsic  exponential distribution of EWs (exponential scale $W_0$ = 30, 50, 150, 300\AA). The panels  from the top to the bottom show \lya\ luminosity distributions, ratios of \lya\ luminosity distributions to the input ones,  EW distribution, and the ratios of  EW distribution to input ones. See Section 4 {\color{black}and 5} for details.}
\label{ewfracevolve}
\end{center}
\end{figure*}

In the upper panel of Figure \ref{ewfracevolve} the \lya\ luminosity distributions of the input samples are plotted as grey lines for various $W_0$, which is simply a Schechter Function. We also plot the ``observed" \lya\ luminosity distribution by adding gaussian errors to expected narrow and broadband flux densities and re-extract their \lya\ luminosities (dark black lines). The ``observed" luminosity distributions are slightly different from the intrinsic ones because of noise fluctuations. 
The dark lines are well consistent with the grey lines at high luminosities, but slightly higher than grey lines at lower luminosities (L$_{Ly\alpha}$ $\sim$ 42.6). This is simply the Eddington bias due to photometry uncertainties. 
The light blue  lines plot the distributions of the intrinsic \lya\ luminosity of the samples after applying the first selection criterion CR1, and the dark blue line the distributions of the ``observed" \lya\ luminosity after apply CR1. Clearly CR1 removes most of the faint LAEs below our detection limit (vertical  dot-dashed line in the upper panel). CR2 further excludes more faint LAEs (light and dark green lines). After applying CR3, however, there are still small fraction of faint LAEs with intrinsic \lya\ luminosity below our detection limit could pass the selection criteria, due to noise fluctuations. However, their ``observed" \lya\ luminosities are all above the detection limit (dark red lines). 

We also plot the ratios of the luminosity distributions to the intrinsic ones, to demonstrate the differences between the input LFs (grey lines, ratio = 1) and the output ones for various exponential scale $W_0$ in Figure \ref{ewfracevolve}. We find that the output LFs are generally consistent with the input ones above the detection limit, except for 1) at high $W_0$, the output ``observed" LFs (dark red lines) are consistent with the input ones at high luminosity, slightly higher than input ones (grey lines, ratio = 1) at low to intermediate luminosities due to Eddington bias, and drop only near the detection limit; 2) at low $W_0$ = 50 \AA, the output  ``observed" LFs (dark red lines) are slightly lower  than the input ones (grey lines, ratio = 1, by a factor of $\sim$ 10\%), because the selections exclude sources with EW $<$ 9.0 \AA, which make more contribution to the whole population (EW $>$ 1 \AA\ in the simulations) at smaller $W_0$. Thus Figure \ref{ewfracevolve} shows that the detection and selection processes only produce weak bias to the luminosity function.

We further examine this issue through directly fitting the the luminosity distributions of the simulated samples, as we did to the real LAE sample in Paper {\sc I}. To measure the L$_*$ and $\Phi_*$ of the artificial samples, we first scale the  input  samples (grey lines in Fig. 3) to match the real one, to ensure the number of simulated LAEs  with \lya\ luminosity in range of log$_{10}$(L) $\sim$ 42.6 -- 43.3 and EW $>$ 9.0 \AA\ meet our observational data.
The same scaling factor was then applied to the simulated output LAE samples.
During the fitting we adopt the same luminosity range  log$_{10}$(L) $\sim$ 42.6 -- 43.3 and the same luminosity bins as in Paper {\sc I}.
We also add Poisson noises to the number of sources in each luminosity bin, to simulate the uncertainties of the Luminosity Function.
We find that the output L$^*$ and $\Phi^*$ for the ``observed" samples at different EW$_0$ are generally consistent with the input values within 1 sigma error bars, also suggests only weak bias was introduced to LF by the detection and selection processes. 
At lower W$_0$ $<$ 100 \AA\ slightly higher $\Phi_*$ was obtained. This is because for EW distribution with lower W$_0$, relatively more simulated LAEs with intrinsic EW $<$ 9.0 \AA\ could be selected due to fluctuations.

Since the observed line EWs suffer from negative values, and very large uncertainties (due to the broad-band weak or non-detection), instead of fitting the line EW distribution, we choose to compare the distribution of R to narrow band flux density ratio of the real LAE sample with artificial samples.  Note that the R to narrowband flux density ratio
has a better behaved error distribution (see Malhotra \& Rhoads 2002 and Wang et al. 2009).


In Figure \ref{ratioer} we plot the distribution of EIS-R to narrow band flux density ratio for our real LAEs. Similar to Wang et al. (2009), we find consistent distributions for all our LAE candidates and for the subset of spectroscopically confirmed sources. Through fitting the observed distribution of flux density ratio with artificial samples, we obtain a best-fit intrinsic exponential scale of  $W_0$ = 262$^{+115}_{-34}$ \AA.  We find that the C-statistic (Cash 1979) is better than $\chi^2$ in fitting the distribution, as some of the observed data bins had few counts.  Our best fit $W_0$ has  $C$-$stat/dof$ = 18.2/19.   Figure \ref{fitstat} plot the fitting statistics (both $\chi^2$ and the C-statistic) as a function of $W_0$ or $\sigma_g$.\footnote{During the fitting we excluded one  object with
 f(R)/f(NB) $\sim$ -0.5, because its continuum flux is affected by over-subtraction of a very bright nearby source.}

Our candidates in CDF-S were selected by Finkelstein et al. (2009), based on EIS-B, EIS-R and narrow bands images. We later obtained public MUSYC B, V, R, I, and z band data in the same field though covering a smaller area (Gawiser et al. 2006). We note that while MUSYC-B band is slightly shallower than EIS-B, the MUSYC-R band  image (5-$\sigma$ limit of $\sim$ 0.10 uJy, m(R)$_{lim}$ = 26.5) is considerably deeper than EIS-R band (5-$\sigma$ limit of $\sim$ 0.22 uJy, m(R)$_{lim}$ = 25.6). To utilize this deeper R band image which could put better constraints on line EW measurements, we extracted MUSYC-R flux for all our LAE candidates located in the MUSYC area. In Figure \ref{ratioer}  we also plot the distribution of MUSYC-R to narrow band flux density ratios for our sample. Obviously, with a deeper R band image, we will get fewer "negative" EW values. We only take into account the f(MUSYC-R)/f(NB) range of [-0.025, 0.3] in the fitting, since one source with f(MUSYC-R)/f(NB) $<$ -0.1 is a spurious measurement caused by a bright nearby object, and the spectroscopic success fraction at f(MUSYC-R)/f(NB)  $>$ 0.3 is very low (only 1 is confirmed out of 7 targets).   Fitting to the distribution yields a better constrained exponential scale of $W_0$ = 167$^{+44}_{-19}$ \AA\ with $C$-$stat$/dof = 9.8/12.
If the EW follows a gaussian distribution, we derive a gaussian scale of $\sigma_g$ = 160$^{+43}_{-12}$ \AA\ (Figure \ref{fitstat}).   Both the exponential distribution and gaussian distribution agrees well with the EW distribution. 
Below we take $W_0$ = 167$^{+44}_{-19}$ \AA\ (or $\sigma_g$ = 160$^{+43}_{-12}$ \AA) as our best measurements, as the MUSYC-R data is much deeper than the EIS-R data.

\begin{figure}
\begin{center}
\includegraphics[totalheight=0.4\textheight]{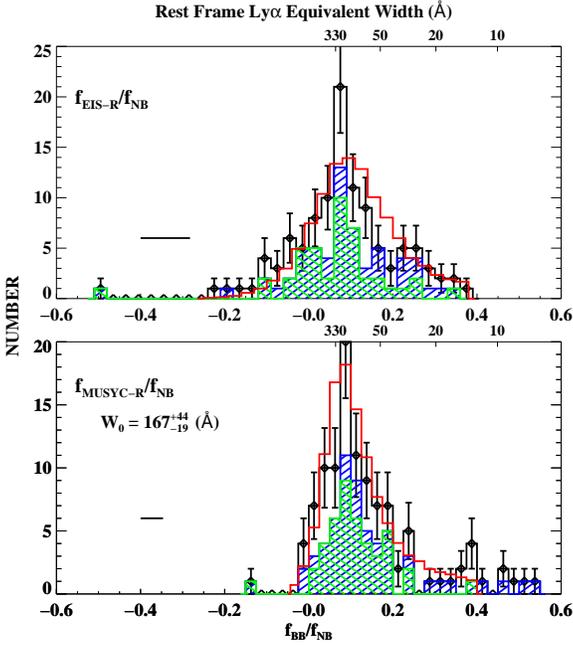}  
\caption{Histogram of the broadband (top: EIS-R band and bottom: MUSYC-R band) to narrowband flux density ratio for our z $\sim$ 4.5 LAE sample. The rest-frame \lya\ line EW (IGM-corrected \lya\ line, marked on the top of the plot) is a monotonic decreasing function of the flux density ratio. The black, blue and green histograms plot distributions for the photometric, targeted, and spectroscopically confirmed samples, respectively. The red line presents the best fit artificial sample assuming the line EW distribution follows an exponential law $d$N$/d$EW = N $exp^{-EW/W_0}$.
The best-fit  EW exponential scale is $W_0$ = 
167$^{+44}_{-19}$ \AA\ from MUSYC-R data alone. The horizontal lines are corresponding to the one sigma error on R band flux divided by the minimum narrowband flux. }
\label{ratioer}
\end{center}
\end{figure}

\begin{figure}
\begin{center}
\includegraphics[totalheight=0.3\textheight]{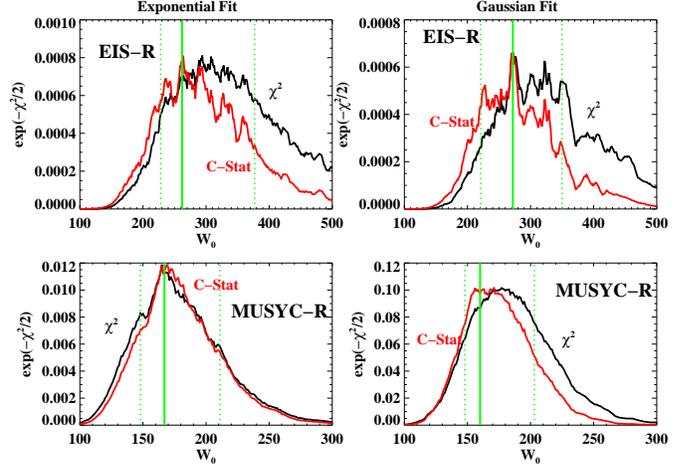}  
\caption{ Statistical probability (P $\propto$ exp(-$\chi^2/2$) for fitting band-ratio distribution with exponential form (left column)  and Gaussian form (right column) of intrinsic EW distribution, with MUSYC-R data on bottom and EIS-R data on top. We compare the statistics with C-statistics (red lines) and $\chi^2$ statistics (black lines), and find that C-statistics give better constraints in the fitting. The green vertical lines are the best fitted values with C-statistics, and their $\pm$1$\sigma$ range are plotted with green dotted lines.  }
\label{fitstat}
\end{center}
\end{figure}

\section {DISCUSSION}
\label{sec:diss}

\subsection{The Selection Effect on the EW distrbution}
\label{sec:diss:select}

The intrinsic exponential scale $W_0$ we obtained from Monte-Carlo simulations is much larger than that from direct fitting to the observed EW distribution (see \S 3.1).
An exponential scale of $W_0$ = 167$^{+44}_{-19}$ \AA\ implies that  42\% (by integrating the exponential  form) of EW $>$ 9.0 \AA\ objects should have EW $\geq$ 154\AA, in good agreement with the fraction in our real LAE sample (39\% of all candidates, and 44\% of spectroscopically confirmed LAEs).
This indicates that the selection bias caused by our selection procedures to the EW distribution is rather weak, if there is any.


To investigate the possible selection bias in detail, we plot in Fig. \ref{ewfracevolve} the EW distributions of the input and output samples of our simulations. 
For output samples, due to noise fluctuations, many of the ``observed" EWs are ``negative" or very high with large errors, similar to the real LAE sample (see \S3). In this figure, we plot the intrinsic EW for the output samples to check the selection effects on sources with different intrinsic EWs.
We find that the recovery rate remains almost constant at EW $>$9.0 \AA, except for the trend that the
recovery rate slightly increases with decreasing EW from $\sim$ 100 to  $\sim$ 20 \AA, and drops only very close the cutoff.
We note that a significant fraction of sources with EW below the cutoff could also survive the selection criteria, due to the noise fluctuations in narrow- and broad-band photometry.  Therefore, Figure \ref{ewfracevolve} presents a quasi Eddington bias pattern that the selection processes yield slightly higher recovery rate for  sources with EW $<$ 154 \AA\ than those with EW $>$ 154 \AA. This is because that for sources with lower line EWs, the contribution from the continuum emission boosts the narrowband flux density, yielding more detections in narrowband, and some low EW objects with EW$<$ 9.0 \AA\ get into the sample.

\subsection{The Evolution of EW Distribution}
\label{sec:diss:evolution}
\begin{figure}
\begin{center}
\includegraphics[totalheight=0.3\textheight]{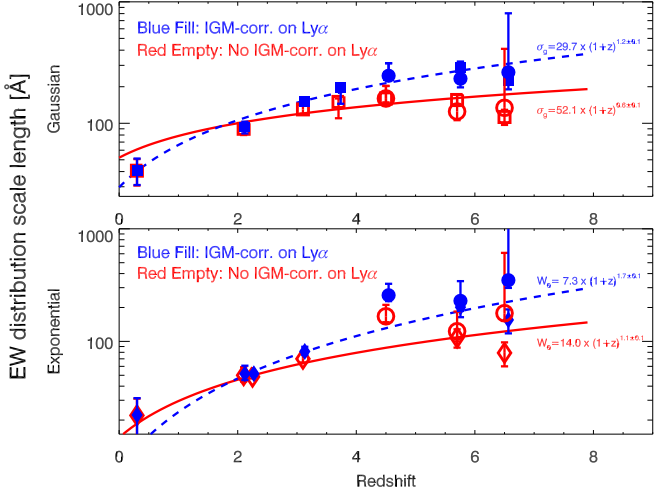}
\caption{ The EW distribution scale {\color{black}(Bottom: exponential distribution scale $W_0$; Top: Gaussian distribution scale $\sigma_g$) obtained at different redshifts. The diamonds and squares present direct fitting results by assuming exponential and gaussian distributions, respectively, and circles present our simulation approach with two kind of distributions. We also mark results with/without IGM absorption-correction to \lya\ line flux in blue/red colors, and we use IGM absorption-correction by Madau (1995) assuming intrinsic symmetric \lya\ emission line with zero velocity offset.} Data references: Cowie et al. (2011) at z $\sim$ 0.3,  Ciardullo et al. (2012) at z = 2.1 and z = 3.1; Nilsson et al. (2009) at z = 2.25 ;Ouchi et al. (2008) at z = 3.1, and z = 3.7,  this work at z = 4.5, Kashikawa et al. (2011) at z=5.7 and 6.5, and Hu et al. (2010) at z = 5.7 and 6.5. {\color{black}The results at z $\sim$ 0.3 are from EGS field sample after AGN have been excluded. }
}
\label{ewevol}
\end{center}
\end{figure}

Does W$_0$  evolve with redshift? 
Below we present a comparison of the EW distributions at various redshifts.
We have shown through simulations that our LAE selection procedures only produce weak bias to EW distributions (Fig. \ref{ewfracevolve}), 
and the major cause of the difference in W$_0$ from direct fitting to the EW distribution (W$_0$ = 56\AA, Fig. \ref{ewrlow}) and from simulations (W$_0$ = 167\AA) is the LAEs with extremely large or even ``negative" EWs.
Therefore our W$_0$ derived through simulations can be compared with measurements in other works through fitting the EW distributions directly,
as long as the underlying broadband images are deep enough to give better constraints on EWs,
or those LAEs with extremely large or even ``negative" EWs have been accounted for correctly.
We note that requiring a broadband detection of LAEs during source selection would also produce severe bias in the LAE EW distribution, since such an approach would naturally exclude sources with large EWs.

For low and moderate-redshift LAEs, spectroscopic observations are easy and powerful to exclude interlopers and AGN, and the underlying broadband images are often deep enough to put good constraints on EW measurements. After excluding AGNs found by Finkelstein et al. (2009b) in the z$\sim$ 0.3 LAEs (Cowie et al. 2010, EGS data only), we get an EW scale of W$_0$ = 22$\pm$9 \AA\ {\color{black}($\sigma_g$ = 41$\pm$10 \AA)}.  This is significantly lower than that of 75\AA\ from fitting the whole photometric sample.
Ciardullo et al. (2012) reported EW scale length of W$_0$  = 50$^{+9}_{-6}$ \AA\ at z=2.1 and 70$^{+7}_{-5}$ \AA\ at z=3.1. Similar or even higher values were also reported by different works, including $W_0$ = 48.5$\pm$1.7 \AA\ at z= 2.3 (Nilsson et al. 2009), $\sigma_{g}$ = 130$\pm$10 \AA\ at  z = 3.1 and $\sigma_{g}$ = 150$^{+10}_{-40}$ \AA\ at  z = 3.7 (Ouchi et al. 2008). 
With the spectroscopically confirmed LAEs at z = 5.7 and 6.5 from Kashikawa et al. (2011), we get a direct fitting EW scale of W$_0$ = 108$\pm$20 \AA\ and 79$\pm$19 \AA\ {\color{black}($\sigma_g$ = 156$\pm$17 and 113$\pm$16 \AA)}, respectively. By applying our  Monte-Carlo simulation approach,  we also obtained the EW scale for z = 5.7 and 6.5 LAEs from Hu et al. (2010), which are $W_0$ = 123$^{+61}_{-23}$ and 178$^{+433}_{-26}$ \AA\  {\color{black}($\sigma_g$ = 125$^{+47}_{-19}$ and 134$^{+276}_{-5}$ \AA)}, respectively.  


Comparing the EW distribution at various redshifts suggests a strong evolution over redshift range of 0.3 to 6.5 (see Fig. \ref{ewevol} \deleteline{}{, blue symbols}). Note in the figure the red data points are without corrections to IGM absorption on the \lya\ line, which itself clearly evolves with redshift too.
Assuming the \lya\ line is symmetric and with no velocity offset from the rest frame of the galaxies, we apply IGM absorption corrections (Madau 1995)
to $W_0$ (blue data points).
Note IGM correction could be different in case of shifted \lya\ lines respected to its rest-frame as seen in observations at high redshifts (e.g. Hashimoto et al. 2013). In order to quantify the evolution we fit the data-points with an analytical function $W_0(z)$  = $A\times(1+z)^\xi$.  We obtain  $A$ = 14$^{+2.8}_{-2.3}$ and $\xi$ = 1.1$\pm$0.1 for {\color{black}exponential scale} $W_0$ before corrections for IGM absorption of the \lya\ line,  and $A$=7.3$^{+1.5}_{-1.3}$ and $\xi$ = 1.7$\pm$0.2 after. {\color{black}With gaussian distribution assumption, the scale values of $\sigma_g$ increase, while the redshift evolution slopes flatten ($A$ = 52$^{+10}_{-8}$ and $\xi$ = 0.6$\pm$0.1 before IGM-correction, and $A$ = 30$^{+6}_{-5}$ and $\xi$ = 1.2$\pm$0.1 after).}

Recent infrared spectroscopy show that there are velocity offsets between rest-frame optical lines compared to \lya\ line peak, which may tell the existence of outflows with velocities of hundreds km/s (Mclinden et al. 2011, Finkelstein et al. 2011).  If the line is offset to the red of systemic velocity,  IGM correction to \lya\  flux is not needed. However, Hashimoto et al. (2013) also reported one LAE with $\sim$0 velocity offset, and an inverse correlation between velocity offsets and \lya\ EWs for LAEs at z$\sim$ 2--3 (see their figure 7, note Shapely et al. 2003 reported similar relation from the composed spectra of z $\sim$ 3 LBGs). This implies that the escape of \lya\ photons could be more complex than a simple outflow enhanced escape. Likely, IGM corrections to \lya\ fluxes need to be applied when there is no velocity offset  and not applied for LAEs with velocity offsets of hundreds of km/s.

\begin{figure}
\begin{center}
\includegraphics[totalheight=0.35\textheight]{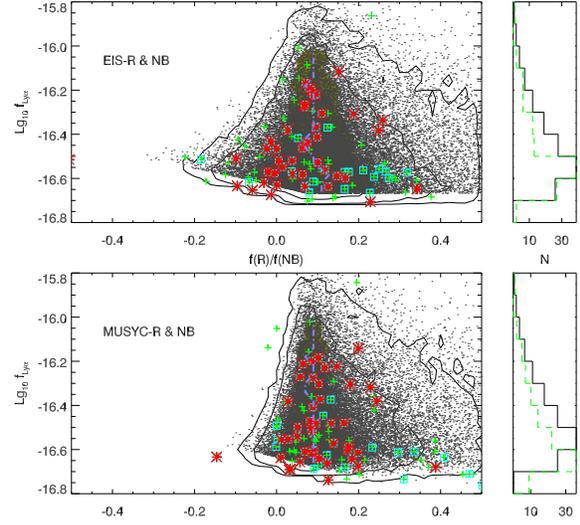}
\caption{Two dimensional distribution of \lya\ line flux and broadband-over-narrowband ratio for our observational LAEs and simulated sources. Green plus and cyan squares are the candidates and targets, and red stars are the confirmed LAEs (Bottom: EIS-R \& NB data; Top: MUSYC-R \& NB data, one confirmed LAEs located outside the simulation region is due to the bright neighbor object in the R-band image). The grey points are the simulations with one parameter of exponential scale W$_0$ = {\color{black}167} \AA\ on both images, black contours show 50\%, 90\%, and 99\% included regions, and  the  distribution of the peak value and FWHM of  band-ratio distributions as a function of \lya\ flux in simulation are plotted as the dashed blue line and the dark yellow region. On the right panel  the distributions of  \lya\ flux for observation (green dashed, all candidates) and simulation (black solid) are shown for EIS-R \& NB data (bottom) and MUSYC-R \& NB data (top), respectively. }
\label{ratioflux}
\end{center}
\end{figure}

\subsection{Is EW$_{Ly\alpha}$ independent of L$_{Ly\alpha}$?}

Through Monte Carlo simulations we have measured the intrinsic EW distribution of our LAE sample at z = 4.5, which is fitted with an exponential law with $W_0$  = 167$^{+44}_{-19}$ \AA.  During our simulation we have assumed that the Ly$\alpha$  EW distribution is independent of Ly$\alpha$  luminosity. However, we do not know prior to this work whether this assumption is correct.

In Figure \ref{ratioflux} we plot the Ly$\alpha$ line flux versus R to narrow band flux density ratio for our LAEs. Since the R to narrow band flux density ratio is a good indicator of the line EW, such a figure provides an opportunity to examine whether the EW distribution is independent to Ly$\alpha$ line luminosity. 
In the figure we see no clear correlation between the Ly$\alpha$ line flux and the
line EW (R to narrow band flux density ratio), but larger scatter in R to narrow band flux density ratio at lower Ly$\alpha$ fluxes.
In Figure \ref{ratioflux} we over-plot the contour distributions of our simulated LAE sample (with $W_0$ = 167\AA). The peak and FWHM of  the distribution of the R to narrow band flux density ratio  from simulated LAE sample at different \lya\ flux bins are also plotted. We see a similar trend in the simulated samples, that LAEs tend to have constant peak R to narrow band flux density ratio  (therefore EW) at different \lya\ line fluxes.  A two dimensional  Kolmogorov-Smirnov test shows that the distributions of the Ly$\alpha$ line flux versus R to narrow band flux density ratio are indistinguishable between the real LAE sample and the artificial sample. This indicates that the intrinsic Ly$\alpha$ EW distribution of our LAE sample at z $\sim$ 4.5 is consistent with our assumption that it is independent to Ly$\alpha$ luminosity.

\begin{figure}
\begin{center}
\includegraphics[totalheight=0.25\textheight]{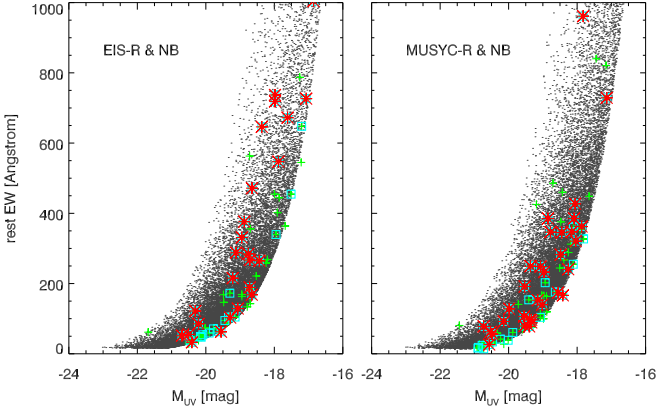} 
 \caption{The rest \lya\ EW vs. UV magnitude (the "Ando" effect) for our sample compared with our simulation for EIS-R \& NB (left) and MUSYC-R \& NB (right). Green plus and cyan squares are the candidates and targets, and red stars are the confirmed LAEs. Note that in this plot, we ignore the LAEs with continuum non-detections.   } 
\label{ando}
\end{center}
\end{figure}

\subsection{``Ando" Effect}
In Figure \ref{ando} we plot the rest frame Ly$\alpha$ EW versus UV magnitude for our LAEs. 
Here the UV magnitude refers to the continuum emission in R band excluding the contribution from the \lya\ line, and applied a correction factor for IGM absorption (see \S3). 
Here, we do not plot those LAEs with ``negative" line EW are ignored, as their negative continuum flux
measurements yield undefined UV magnitudes.
In such a plot we see a clear lack of large EW LAEs with large UV luminosities, and the maximum LAE EW in
the sample systematically decreases with increasing UV luminosity. This effect is known as the ``Ando effect,'' as first reported in LBGs at z $\sim$ 5-6 by Ando et al. (2006), and also detected in LAEs by later works (Shimasaku et al. 2006, Stanway et al. 2007, Deharveng et al. 2008, Ouchi et al. 2008). However, by over-plotting the artificial LAE samples we have simulated, we show that this effect in our sample could be naturally generated through our LAE selection.
Actually, the $EW_{Ly\alpha}-M_{UV}$ plane can be expressed as $EW_{Ly\alpha}-L_{Ly\alpha}/EW_{Ly\alpha}$ plane, which is an inverse relation as seen in Figure \ref{ando}.


\subsection{Implication for NB selection}

As we have stated previously, the selection of LAEs relies on both the depth of the narrow-band and the underlying broad-band images. The simulation procedures we have developed provide a powerful approach to test the selection efficiency under various conditions, and such tests could be used to guide  future narrowband imaging surveys.

We adopt one quantity to describe the efficiency of selections for our simulated samples: the number of LAE selected. 
In Figure \ref{nbsel} we plot the results of our simulations for various given conditions. We clearly see that deeper narrowband images yield more LAE candidates, but the role of the broadband image depth is also important. If the limiting magnitude of the broadband image is shallower than the narrowband image, the selection efficiency is poor in the number of LAE selected, and large errors in broad band image will introduce larger uncertainties in the EW calculation. The selection efficiency steadily rises with deeper broadband image. However, the rise slows down or even halts if the broadband images are $>$ 0.5 -- 1.0 mag deeper than the narrowband, indicating that broadband images much deeper than the narrowband will not increase the number of LAEs selected if one keeps the selection criteria.
Actually, for selections with much deeper broadband images, the proper approach is to go for line emitter with smaller EW, thus could further increases the number of sources selected. Therefore, deeper broadband images would be always helpful in selections of line emitters, but broadband images significantly shallower than the narrowband would be very inefficient. Similar patters could be seen for different $W_0$ (Figure \ref{nbsel}).



\begin{figure}
\begin{center}
\includegraphics[totalheight=0.3\textheight]{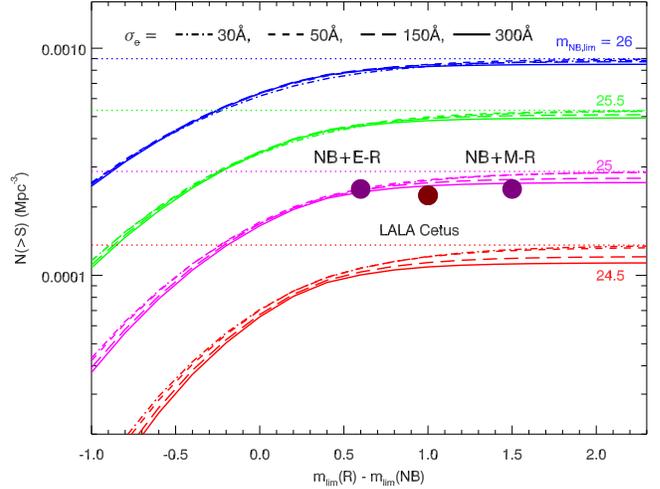}  
\caption{Cumulated number density as a 
function of R band and narrowband limits (5 $\sigma$ limit). The narrowband depth are marked in 4 steps from m$_{lim}$(NB) = 24.5 to 26 presented by different colors.   Simulation with $W_0$ = 30, 50, 150, and 300 \AA\ are presented as dash-dot, dash, long-dash and solid lines, respectively. The purple  filled circles show the narrowband and broad band depth for our narrowband images with EIS-R or MUSYC-R band data, and the brown filled circle shows the depth of LALA Cetus field (Wang et al. 2009).   }
\label{nbsel}
\end{center}
\end{figure}



\section{Summary}
In this work we study the intrinsic \lya\  EW distribution of our z$\sim$ 4.5 LAEs in ECDFS. 
To derive the intrinsic line EW distribution, we develop essential Monte-Carlo simulations to address the selection effects of our LAE selection procedures, 
and the large uncertainties in line EWs from narrow- and broad-band photometry. 

Our approach includes 1) build artificial LAE samples following given \lya\ luminosity function and EW distribution; 
2) add observational uncertainties to their expected narrowband and the underlying broad photometry;
3) run our LAE selection processes to recover the simulated LAEs; 
and 4) compare the simulated LAE sample with the real LAE sample we obtained in ECDFS, specifically compare their
luminosity function and EW distribution. We note the comparison of EW distribution is performed on the
distributions of the narrow- to broad-band flux ratio between simulated and real samples, since the narrow- to broad-band flux ratio
 is a monotonic decreasing function of line EW and has a much better behaved error distribution. 
\\
\\
Our main results are summarized as below:
\begin{itemize}
\item With simulations, we find that our LAE selection procedures produce weak (quasi-) Eddington bias to both \lya\ luminosity function and EW distribution.
\item Direct fitting on EW distribution gives an exponential scale of W$_0$ = {\color{black}50}$\pm$11 \AA, while after taking into account the broadband non-detections,  we get W$_0$ = 167$^{+44}_{-19}$\AA\ (or gaussian scale of $\sigma_{EW}$ = 160$^{+43}_{-12}$ \AA\ ) for LAEs at z$\sim$ 4.5 through fitting on band-ratio distribution .  
\item We find our LAE sample is consistent with an assumption that the intrinsic LAE EW distribution is independent to \lya\ luminosity, which
could naturally produce "Ando" effect in LAE samples. 
\item Our simulations also show that broad band image $\sim$0.5--1 mag deeper than the inside narrowband is most efficient in selecting
emission line sources adopting our selection criteria. The simulations are useful to optimize future similar surveys at various redshifts.
\item We find a strong evolution of the \lya\ EW distribution over redshift 0.3 to 6.5, which can be well fitted by a power-law form $W_0$ $\propto$ (1+z)$^{\xi}$, with $\xi$ = {\color{black} 1.1$\pm$0.1(or $\xi$ = 1.7$\pm$0.2} after applying an IGM-absorption correction to \lya\ line) for EW exponential distribution, {\color{black} and $\xi$ = 0.6$\pm$0.1 (or $\xi$ = 1.2$\pm$0.1 after applying an IGM-absorption correction to \lya\ line) for a gaussian EW distribution}. 

\end{itemize}

\section*{Acknowledgements}
We would like to thank the support of NSF grant AST-0808165 and NOAO TSIP program.  ZYZ would like to thank the funding from the SESE Exploration Postdoctoral Fellowship Program.
The work of JXW is supported by Chinese National Science Foundation through Grant 10825312 \& 11233002. We would like to thank the referee for his careful reading through our paper, and thank Zheng Zheng for his helpful comments on this project.  

\bibliographystyle{mn2e}

\label{lastpage}

\end{document}